# Interdependence of Transmission Branch Parameters on the Voltage Levels

Mir Hadi Athari
Electrical and Computer Engineering
Virginia Commonwealth University
atharih@vcu.edu

Zhifang Wang
Electrical and Computer Engineering
Virginia Commonwealth University
zfwang@vcu.edu

## Abstract

*Transformers and transmission lines are critical components of a grid network. This paper analyzes the statistical properties of the electrical parameters of transmission branches and especially examines their interdependence on the voltage levels. Some interesting findings include: (a) with appropriate conversion of MVA rating, a transformer's per unit reactance exhibits consistent statistical pattern independent of voltage levels and capacity; (b) the distributed reactance (ohms/km) of transmission lines also has some consistent patterns regardless of voltage levels; (c) other parameters such as the branch resistance, the MVA ratings, the transmission line length, etc, manifest strong interdependence on the voltage levels which can be approximated by a power function with different power constants. The results will be useful in both creation of synthetic power grid test cases and validation of existing grid models.*

## 1. Introduction

Modern power systems use multiple voltage levels to decrease energy loss in the transmission network [1]. The voltage level is changed through the extensive use of transmission transformers to step up the voltage for long-distance transmission lines and then step down to lower voltages to go through the distribution network. This multi voltage-level structure causes different grid components to have voltage dependent parameters and features. Branches in power networks are among those components that can have a heavy dependence on voltage level. Generally, in power systems, the term "branches" refers to transmission lines or transformers between two buses in a network. The study of the interdependence of transmission branch parameters on voltage levels can provide useful insights as well as multiple validation metrics for synthetic power networks.

Synthetic power networks are introduced as a potential solution for the restricted access to real-world power system test cases. Confidentiality requirements limit the access to real data in critical infrastructures like power systems. On the other hand, researchers in power industry need realistic test cases of varying sizes and complexities and appropriate properties in order to evaluate and verify their proposed solutions and novel approaches. For example, the algorithms introduced by authors in [2]–[4] need some verification in larger systems to identify the pros and cons of the solution. Another example is the concept of real-time optimal power flow in [5] that can be evaluated in numerous synthetic grids. Since Synthetic power networks are entirely fictitious but with the same characteristics as real networks, they can be freely published to the public to facilitate advancement of new technologies in power systems. One such characteristic is the interdependence of different branch parameters on voltage level.

In the literature, many studies are dedicated for characterizing actual power networks and/or developing a synthetic one, mainly from topological perspectives such as ring-structured power grid developed in [6] and tree structured power grid model to address the power system robustness [7], [8]. Works of [9]–[11] used the small world approach described in [12] as a reference to generate some synthetic transmission network topologies. The *RT-nestedSmallWorld* random topology model proposed in [10] is based on comprehensive studies on the electrical topology of some real-world power grids. Authors in [13] studied the impacts of randomized and correlated siting of generation and loads in a grid on its vulnerability to cascading failures. [14]–[17] defined a topology measure called "bus type entropy" to characterize the correlated siting of generation and load in actual power grids, based on which an optimization algorithm was developed to determine appropriate bus type assignments in a synthetic grid modeling. [18] studied the statistics of generation size and load settings. [19] gave a comprehensive report about the scaling property of power grid in terms of selected topology measures and electric parameters. Authors in [20] reported some initial study results on the statistics of transmission line parameters. The substation placement method and transmission lines assignment based on population and energy data in [21] uses the methodology introduced in [22], [23], where they employ a clustering technique to ensure that synthetic substations meet realistic proportions of load and generation. [24] addressed the need for synthetic large-scale system dynamic models for transient stability studies such as wide-area damping control in [25], [26] and dynamic control allocation for damping of inter-area oscillations in [27]. The collaboration of researchers from five universities has resulted in publishing three fully synthetic power networks called ACTIVSg200, 500, and 2000 cases [21], [24]. Later, they published a set of topological and electrical validation metrics in [28] to assess the realism of the developed synthetic power grids. The authors will continue to augment those cases by adding additional complexities and verification and tuning of the parameters.

Statistical studies on the database from historical weather data for forecasting in [29] to probabilistic methods for reliability assessment based on historical data in [30] and a data-driven analysis on capacitor bank operation in [31] show that statistics derived from real-world data are commonly used for modeling and validation in power systems. The above literature review on synthetic grid modeling suggests that there is a need for a comprehensive statistical study on real-world power systems branch electrical and non-electrical parameters. This will allow us to identify the interdependencies of various electrical and topological parameters on the nominal voltage level. Also, it may provide us with useful guidelines on their distribution to be used in parameter value assignment in synthetic cases. In this paper, using a large sample of real-world power system branch data from Federal Electricity Regulatory Commission (FERC), we present a statistical study to characterize electrical and non-electrical parameters of the transmission network to be used in synthetic grid models. The goal of this paper is: (a) to identify the interdependence of branch parameters on the nominal voltage level and (b) to provide guidelines on how to accurately configure them in the synthetic models.

The rest of the paper is organized as follows. Section 2 analyzes and presents branch parameters that are independent of the voltage level. Section 3 discusses the statistics and interdependence of other branch parameters on voltage levels. In Section 4, the validation of three published synthetic grids according to derived statistics will be presented and finally, some concluding remarks and future work will be presented in section 5.

## 2. Voltage independent parameters

In this study, we focused on seven different parameters from two real-world power systems including transformers and transmission lines per unit and distributed reactance, X/R ratio, transformers and transmission lines capacity ($MVA$), and transmission line length ($km$). We found some of these parameters exhibit a strong correlation with voltage level while others show a very trivial dependence on voltage level which can be assumed approximately voltage independent. The latter includes transformers per unit reactance converted to their own MVA base and transmission lines distributed reactance ($\Omega/km$). In this section, the statistics of these parameters will be presented.

### 2.1. Transformer per unit reactance

Per unit system is a common method used in power system analysis to express the system quantities as fractions of a defined base unit quantity. Considering a large number of transformers deployed in the power systems with different voltage levels for their terminals, the use of per unit system is important. Another advantage for this expression is a common engineering practice in which the transformer impedance falls into a narrow numerical range when expressed as per unit fraction of the equipment rating, even if the unit size varies widely. However, in practice, the per unit impedances of power system components are converted to different values using a common system-wide base and then used in power flow or economic power flow calculations. So, the conversion of per unit impedance of each component can be done back and forth from system-wide common base to equipment's own rating and will significantly impact the range of the parameter. This conversion is based on the following formula that depends on the voltage bases for different zones in the system and a predefined unique power base for the entire system ($S_{base}$):

$$Z_{PU}^{New} = Z_{PU}^{Given} \times \left(\frac{V_{Base}^{Given}}{V_{Base}^{New}}\right)^2 \times \left(\frac{S_{Base}^{New}}{S_{Base}^{Given}}\right) \quad (1)$$

where $Z_{PU}^{Given}$, $V_{Base}^{Given}$, $S_{Base}^{Given}$ are given per unit impedance, voltage base, and power base for each apparatus and $Z_{PU}^{New}$ is the new per unit impedance calculated using $V_{Base}^{New}$ and $S_{Base}^{New}$. Usually, the voltage base values are selected the same as the nominal voltage of transformer terminals for each zone to simplify the calculations. Therefore, the conversion formula for per unit impedance is expressed as

$$Z_{PU}^{New} = Z_{PU}^{Given} \times \left(\frac{S_{Base}^{New}}{S_{Base}^{Given}}\right) \quad (2)$$

In this study, the transformers are grouped based on their high voltage side to examine their parameters interdependence on the voltage level. The original data from FERC were reported in per unit values based on the system-wide common base. Our initial observations in [32] show that per unit reactance calculated based on system common base falls into a wide range and we are not able to find a standard probability distribution for them. However, after converting them into values based on transformer own rating, they fall into a narrow range regardless of their size. In other words, there exists no interdependence between per unit reactance and the voltage level of transformers after this conversion as shown in Figure 1.

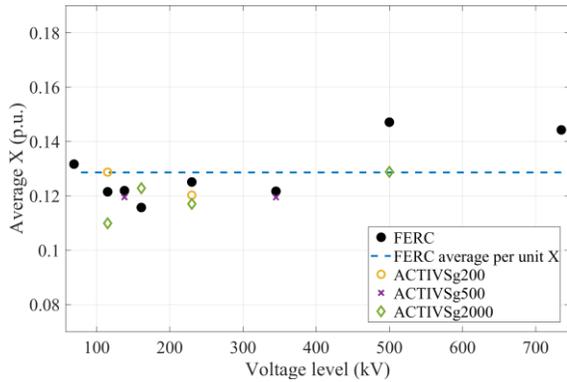

**Figure 1. Interdependence of transformer per unit reactance on voltage level**

In this figure, the black dots are the average per unit reactance of transformers for different voltage levels from 69 to 735 kV. The blue dashed line is the average of all data points. There is no visible trend in the data which means the per unit reactance of transformers calculated based on their own rating is independent of the voltage level.

In addition, we found that there exist some extraordinarily large values for transformer per unit reactance in the original data from FERC. These values make it difficult to fit a standard PDF to data. Furthermore, the range of the data becomes very large while including the outliers. In this study, we remove the outliers from all data points to avoid erroneous disturbance on statistical analysis. The outliers are removed based on box plot method where values beyond a certain threshold are considered *extreme outliers* and exclude when fitting an empirical PDF curve to the data. It is found that excluding outliers from data set leads to the more consistent statistical pattern for the parameters. For example, the Normal distribution found to perfectly fit the transformer per unit X after excluding the outliers while in [32] the t Location-Scale distribution was recognized the best fit to the parameter.

Figure 2 shows the probability distribution of per unit reactance of transformers for four different voltage levels. It is found that this parameter can be approximated using the Normal distribution. The goodness of this fit is measured with Kullback-Leibler divergence. In probability theory and information theory, the Kullback–Leibler (KL) divergence, also called discrimination information, is a measure of the difference between two probability distributions P and Q. It is not symmetric in P and Q. In applications, P typically represents the "true" distribution of data, observations, or a precisely calculated theoretical distribution, while Q typically accounts for a theory, model, description, or approximation of P [33]. Specifically, the KL divergence from Q to P, denoted $D_{KL}(P \parallel Q)$, is the amount of information lost when Q is used to approximate P. For discrete probability distributions P and Q, the KL divergence from Q to P is defined to be [34]

$$D_{KL}(P \parallel Q) = \sum_i P(i) log \frac{P(i)}{Q(i)} \quad (3)$$

In words, it is the expectation of the logarithmic difference between the probabilities P and Q, where the expectation is taken using the probabilities P. Therefore, smaller values for the divergence represents a more accurate fit for the empirical PDF of the parameters.

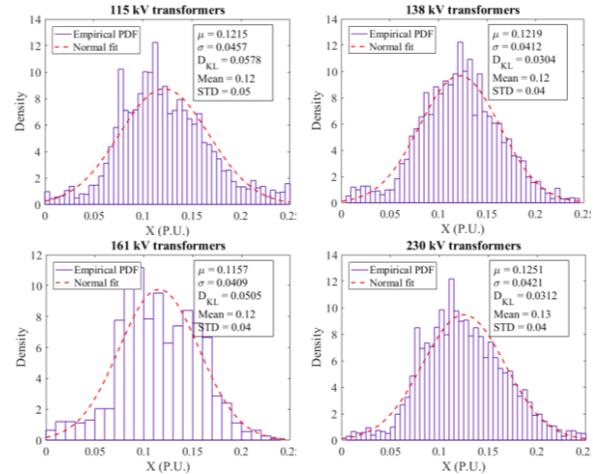

**Figure 2. Empirical PDF and the Normal fit of per unit reactance for 115, 138, 161, and 230 kV transformers**

As shown in figure 2, the per unit reactance in transformer own rating is within a fixed range (0 to 0.25 P.U.) for all voltage levels. Also, they all can be best fit with the Normal distribution with relatively small KL divergence values meaning that we lose a minimal amount of information by using the Normal distribution for this parameter.

## 2.2. Transmission line distributed reactance

Transmission line distributed reactance ($\Omega/km$) is the second parameter that shows no dependence on voltage level. The original data from two real-world power systems are reported in per unit values. In order to convert per unit values into distributed reactance, we use a formulas as follows:

$$X(\Omega/km) = \frac{X_{pu} V_B{}^2}{l\, S_B} \qquad (4)$$

in which using system common base $S_B$ and voltage base $V_B$ for each transmission line the actual reactance in *ohms* is first calculated; then using the approximated line length $l$ in *km*, the distributed reactance in $\Omega/km$ is then derived. Note that, the line length data reported from FERC is approximated and are calculated using Geographical Information System (GIS) data. This may not have a big impact on long lines, while it can affect shorter lines, as the actual distance between two buses may be longer than the direct line between the geographic locations of the two buses. Figure 3 shows the distributed reactance of transmission lines for different voltage levels. The black dots are average distributed reactance for each voltage level and the blue dashed line is their average. Similar to transformer reactance, we can see no visible interdependence between these two parameters which means the distributed reactance of transmission line is an independent parameter from the nominal voltage level.

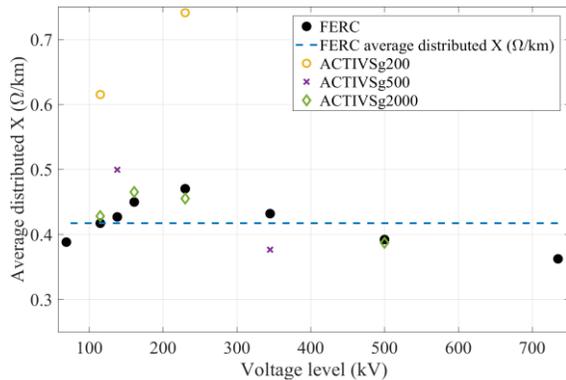

**Figure 3. Interdependence of transmission line distributed reactance on voltage level**

We also examined the distribution of per unit reactance of transmission lines for select voltage levels. Figure 4 shows their distribution and approximated exponential fit using KL divergence criteria. Note that, since we could not find a standard fitting function for distributed reactance ($\Omega/km$), we used per unit reactance instead that shows a clear exponential decay for all four considered voltage levels. However, the mean value of each distribution function also indicates a strong correlation with the voltage levels.

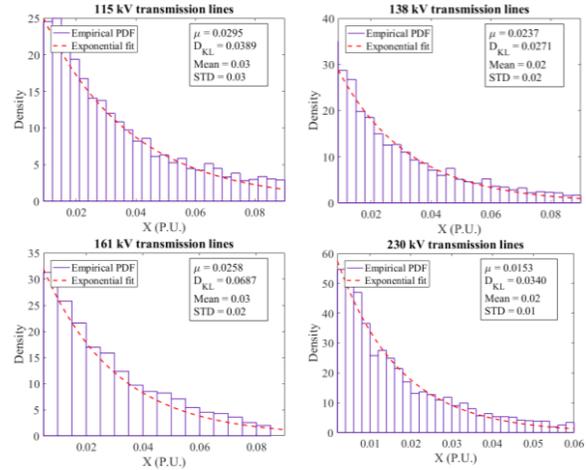

**Figure 4. Empirical PDF and exponential fit of per unit reactance for 115, 138, 161, and 230 kV transmission lines**

## 3. Voltage-dependent parameters

Out of seven studied parameters from network branches, five shows a very strong interdependence on voltage level which can be used in validation and tuning of existing synthetic models such as ACTIVSg cases. Following is the detailed results of analysis on voltage-dependent branch parameters in real-world power networks. To characterize the interdependence of each parameter on nominal voltage level mathematically, using power curve of $(V_B) = a \times V_B{}^b$, their empirical relationship will be extracted. The choice of power function makes it easier to validate the empirical results with physical constraints of the network imposed by Kirchhoff's voltage and current laws and Ohm's law which is the subject of our next study.

### 3.1. Transformer capacity (MVA)

When transformers are grouped based on their high voltage side, there is a visible trend in their size. We found that the larger the voltage level the bigger the transformer size. Figure 5 shows the interdependence of transformer capacity (*MVA*) on voltage levels. In this figure, black dots show the average transformer size for each voltage level and the blue dashed curve, represent a power equation that is fit to these data considering minimum Root Mean Squared Error (RMSE).

According to the curve fitting result for transformer capacity versus voltage level, the transformer capacity in MVA is related to its voltage level in the form of $S_{TX} = 0.172 . V^{1.332}$. This can be served as a validation and tuning metric to adjust the size of transformers in the synthetic grids.

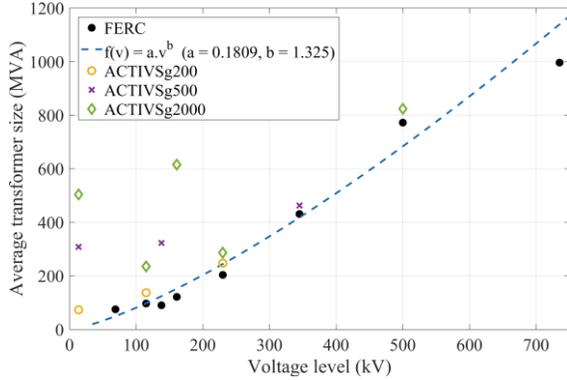

**Figure 5. Interdependence of transformer capacity on voltage level**

Figure 6 shows the distribution of transformer size and an approximated Generalized Extreme Value (GEV) fit for select voltage levels. The Cumulative Density Function (CDF) for GEV distribution is represented by (4)

$$F(x|\zeta,\mu,\sigma) = exp\left(-\left(1 + \zeta\frac{(x-\mu)}{\sigma}\right)^{\frac{-1}{\zeta}}\right) \quad (4)$$

where $\mu$ is location parameter, $\sigma$ is scale parameter, and $\zeta \neq 0$ is shape parameter. Using this mathematical distribution, one can generate reasonable values for transformer capacities in a given synthetic grid model.

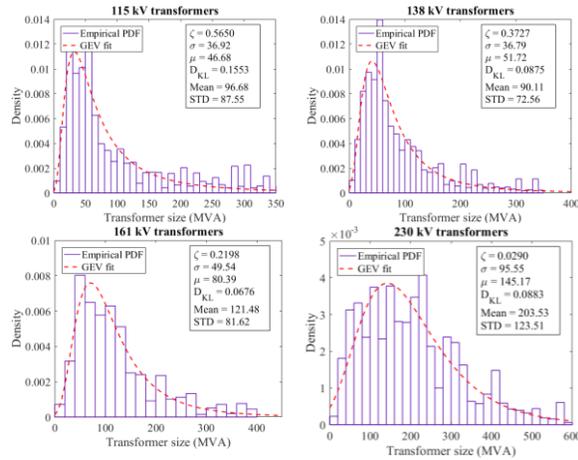

**Figure 6. Empirical PDF and GEV fit of transformer capacity for 115, 138, 161, and 230 kV transformers**

## 3.2. Transformer X/R ratio

Another voltage dependent electrical parameter is the ratio between per unit reactance and per unit resistance of the transformer. Our analyses suggest that as the size of transformer grows, their X/R ratio increases as well (see Figure 7).

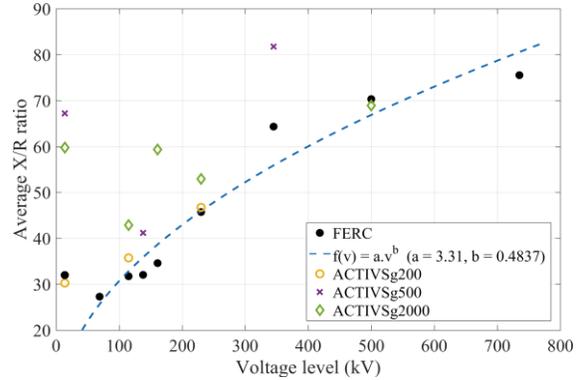

**Figure 7. Interdependence of transformer X/R ratio on voltage level**

Similar to transformer capacity, the relationship between X/R ratio and voltage level can be expressed using power function as shown in Figure 7. This is another metric useful for validation and tuning purposes in synthetic grid modeling.

The empirical distribution of X/R ratio for transformers with different voltage levels and GEV fit are depicted in Figure 8. All distribution fittings show small KL divergence value which is a metric for the goodness of the approximated fit for the data.

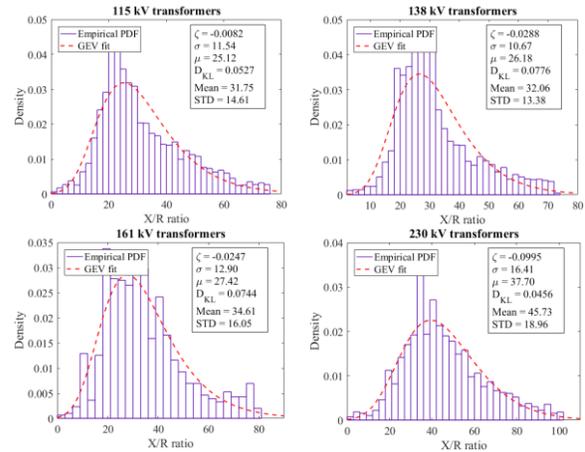

**Figure 8. Empirical PDF and GEV fit of transformer X/R ratio for 115, 138, 161, and 230 kV transformers**

## 3.3. Transmission line length

As mentioned earlier, the length of the transmission lines (*km*) in different voltage levels is calculated based on GIS data and the great circle method. While this approximation may not exactly reflect the line length, the data can be used to examine the interdependence of average line length on voltage level. Figure 9 shows the relationship between average line length and voltage levels for transmission lines of 69 to 735 kV. Using the similar procedure as used in the last two parameters, the curve fitting based on power function is performed and the fitting parameters are shown in the figure.

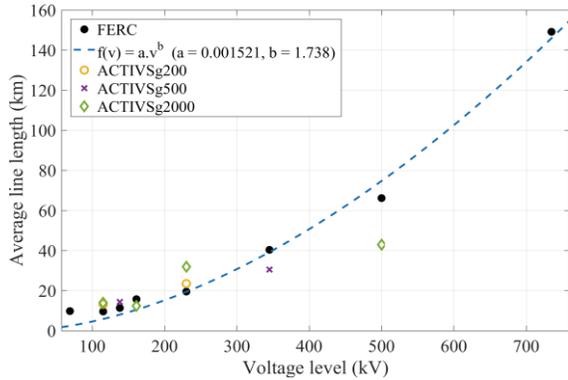

**Figure 9. Interdependence of transmission line length on voltage level**

The distributions of transmission line length and the approximated GEV distribution are shown in Figure 10. It is found that, as the voltage level in transmission lines increases, the average line length grows as well which is consistent with the common engineering practice in power systems. To reduce power loss in long lines of the network, higher voltage levels are used which in turn leads to the reduced current in the line and consequently, the power loss along the line drops significantly.

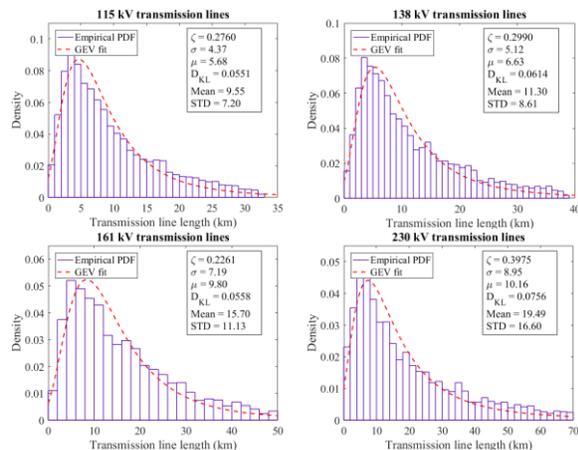

**Figure 10. Empirical PDF and GEV fit of transmission line length for 115, 138, 161, and 230 kV transformers**

### 3.4. Transmission line X/R ratio

Another parameter that we examined for their interdependence on voltage level is the X/R ratio for lines at different voltage levels. This parameter is important for tuning purposes because given valid reactance values for transmission lines, this ratio helps us assign valid values to the line resistance. Figure 11 shows the interdependence of this parameter on voltage level for transmission lines of 69 to 735 kV. We can observe an almost linear increase in the X/R ratio as voltage level increases.

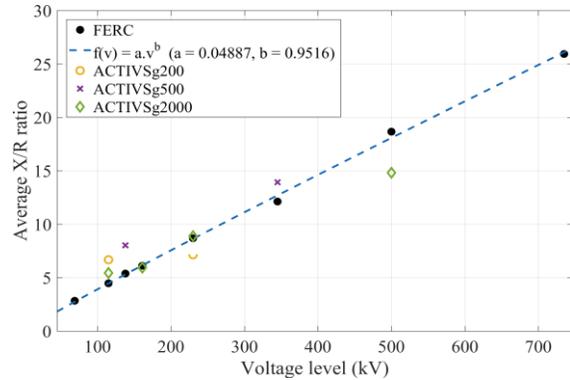

**Figure 11. Interdependence of transmission line X/R ratio on voltage level**

The blue dashed line is fitted to average X/R ratio points using power function. The fitting parameter $b$ is calculated as 0.95 that shows an almost linear relationship between these two parameters.

Finally, the distribution of X/R ratio for transmissions lines is shown in Figure 12. Based on KL divergence, the Normal distribution found to be the best fitting curve as shown in the figure.

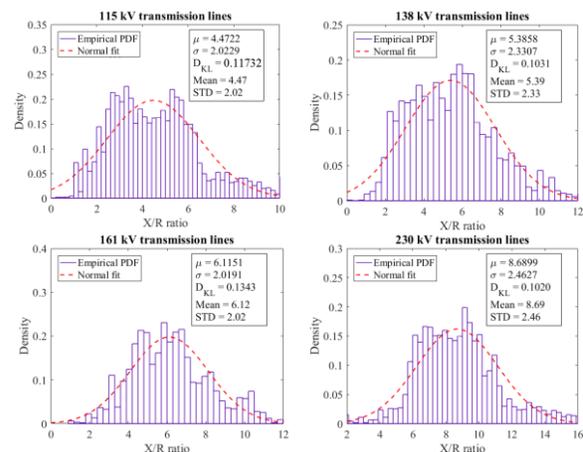

**Figure 12. Empirical PDF and the Normal fit of transmission line X/R ratio for 115, 138, 161, and 230 kV transformers**

### 3.5. Transmission line capacity (MVA)

Similar to transformers, transmission lines on different voltages have different capacities. Figure 13 shows the interdependence of transmission line capacity (MVA) on the nominal voltage level for the FERC data. Black dots show the average line capacity per voltage level and the blue dashed line is calculated based on the curve fitting using power function. The curve fitting parameters are shown in the figure.

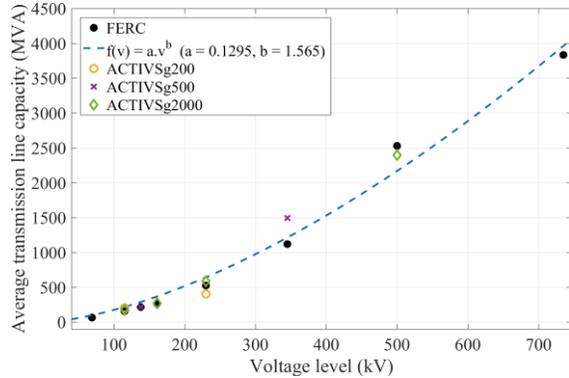

**Figure 13. Interdependence of transmission line capacity on voltage level**

Finally, Figure 14 shows the distribution of line capacity data for different voltage levels. Unlike the capacity of the transformer, the best fitting function found to be the Normal distribution with the minimum KL divergence. Among voltage levels, 161 kV transformers exhibit the largest KL divergence for the Normal fit. However, the Normal distribution function was the best fit to the data.

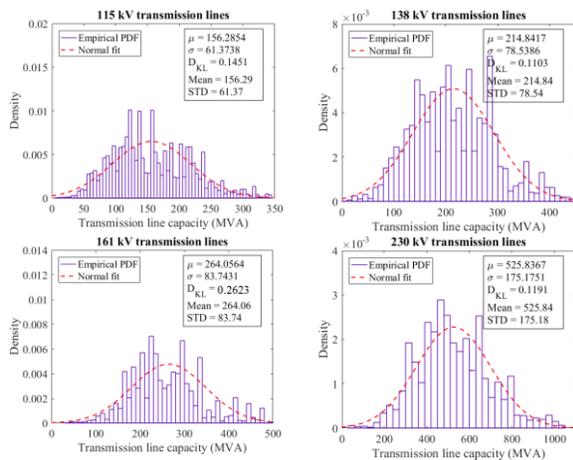

**Figure 14. Empirical PDF and the Normal fit of transmission line capacity for 115, 138, 161, and 230 kV transformers**

### 4. Validation results with some synthetic grid models

In this section, we try to compare and validate the seven parameters for the ACTIVSg cases in terms of their interdependence on the nominal voltage level. For comparison purposes, the average values of each parameter for different voltage levels are superimposed on the figures presented in the previous sections (see Figures 1, 3, 5, 7, 9, and 11).

For transformer per unit reactance (Figure 1), all three ACTIVSg cases are within the scope and present independent values from voltage level which is consistent with what is found from FERC data. For transmission line distributed reactance (Figure 3), ACTIVSg500, and ACTIVSg2000 cases show comparable values with those of FERC data and there is no visible trend in the data, while ACTIVSg200 case seems to have some extraordinarily large values that make the average larger. For transformer capacity (Figure 5), all three cases show an increasing trend with respect to the voltage level which is consistent with the real data from FERC. However, the ACTIVSg500 and 2000 cases seem to have oversized transformers for lower voltage levels. Transformer X/R ratio for ACTIVSg500 and 2000 seems a bit out of order (see Figure 7) and they don't exhibit the same growth trend with regard to the voltage level as we recognized in FERC data. But, the ACTIVSg200 case exhibits a consistent trend with that of actual data. For transmission line length, X/R ratio, and capacity (see Figures 9, 11, and 13), all three ACTIVSg cases exhibit similar trend with close values to those found from statistical analysis on the real data. Table 1 summarizes the validation results for the three synthetic grid cases. Note that in the table, the check mark denotes to the consistency of the parameters and statistics from the real data for the corresponding ACTIVSg case, while we used TR (Tuning Required) for parameters whose average value don't fall within the scope of those found from FERC data.

Based on the above observations for ACTIVSg synthetic power system cases, the majority of parameters for these cases are consistent with statistics derived from the real-world systems. However, some parameters such as transformer size in ACTIVSg500 and 2000 cases need to be tuned in order to conform to the real situation. This can be easily addressed by reassigning the transformer capacities based on empirical PDF identified for the parameter (see Figure 6) using the average value shown in the curve fitting result of Figure 5. Similarly, for the transformers X/R ratio and transmission line distributed reactance, the same tuning procedure based on extracted statistics from

FERC data can be applied to cases with out of scope parameter values. This shows the practical application of the presented statistics in the paper in the context of synthetic power system modeling.

**Table 1. Validation on the interdependence of transmission branch parameter on voltage levels**

| Parameter | Synthetic Grid Models | | |
|---|---|---|---|
| | ACTIVSg 200 | ACTIVSg 500 | ACTIVSg 2000 |
| Transformer X (p.u.) | ✓ | ✓ | ✓ |
| Line X (Ω/km) | TR | ✓ | ✓ |
| Transformer Capacity (MVA) | ✓ | TR | TR |
| Transformer X/R ratio | ✓ | TR | TR |
| Line Length l (km) | ✓ | ✓ | ✓ |
| Line X/R ratio | ✓ | ✓ | ✓ |
| Line Capacity (MVA) | ✓ | ✓ | ✓ |

## 5. Conclusions and future work

The statistical properties of the electrical and non-electrical parameters of transmission branches from two real-world power systems are examined in this study. Seven parameters including transformer per unit reactance, transmission line distributed reactance and line length, transformer and transmission lines X/R ratio, and transformer and transmission line capacity are considered in the statistical analysis. It is found that some parameters exhibit strong interdependence on the nominal voltage level such as X/R ratios, branch capacities, and transmission line length, while others show no dependence on the voltage level like transformer per unit reactance calculated based on their own rating and transmission line distributed reactance ($\Omega/km$). Using the power function, the relationship between parameters and the voltage level is extracted and expressed to serve as validation metric and tuning criteria in synthetic grid modeling. These findings will be helpful in both creation of synthetic power grid test cases and validation of existing grid models.

As the future extension of this study, we want to cover a wide range of electrical and non-electrical parameters of transmission branches to provide a comprehensive validation study for synthetic grid modeling applications. In addition, the verification of the empirical parameter-voltage level relationships based on physical constraints of the power system is of interest.

## 6. Acknowledgement

The work in this study was sponsored in part by DoE through the CERTS initiative and the ARPA-E programs under grant number DE-AR0000714.

## 7. References

[1] J. D. Glover, M. S. Sarma, and T. Overbye, *Power System Analysis & Design, SI Version*. Cengage Learning, 2012.

[2] H. Moradi, A. Abtahi, and M. Esfahanian, "Optimal operation of a multi-source microgrid to achieve cost and emission targets," in *2016 IEEE Power and Energy Conference at Illinois (PECI)*, 2016, pp. 1–6.

[3] H. Moradi, A. Abtahi, and M. Esfahanian, "Optimal energy management of a smart residential combined heat, cooling and power," *Int. J. Tech. Phys. Probl. Eng*, vol. 8, pp. 9–16, 2016.

[4] A. Dehghan Banadaki and A. Feliachi, "Voltage Control Using Eigen Value Decomposition of Fast Decoupled Load Flow Jacobian," in *49th North American Power Symposium (NAPS)*, 2017, pp. 1–6.

[5] E. Mohagheghi, A. Gabash, and P. Li, "A Framework for Real-Time Optimal Power Flow under Wind Energy Penetration," *Energies*, vol. 10, no. 4, p. 535, Apr. 2017.

[6] M. Parashar, J. S. Thorp, and C. E. Seyler, "Continuum Modeling of Electromechanical Dynamics in Large-Scale Power Systems," *IEEE Trans. Circuits Syst. I Regul. Pap.*, vol. 51, no. 9, pp. 1848–1858, Sep. 2004.

[7] B. A. Carreras, V. E. Lynch, I. Dobson, and D. E. Newman, "Critical points and transitions in an electric power transmission model for cascading failure blackouts," *Chaos An Interdiscip. J. Nonlinear Sci.*, vol. 12, no. 4, p. 985, 2002.

[8] M. Rosas-Casals, S. Valverde, and R. V. Sole, "Topological Vulnerability of the European Power Grid Under Errors and Attacks," *Int. J. Bifurc. Chaos*, vol. 17, no. 7, pp. 2465–2475, Jul. 2007.

[9] Z. Wang, R. J. Thomas, and A. Scaglione, "Generating Random Topology Power Grids," in *Proceedings of the 41st Annual Hawaii International Conference on System Sciences (HICSS 2008)*, 2008, pp. 183–183.

[10] Z. Wang, A. Scaglione, and R. J. Thomas, "Generating Statistically Correct Random Topologies for Testing Smart Grid Communication and Control Networks," *IEEE Trans. Smart Grid*, vol. 1, no. 1, pp. 28–39, Jun. 2010.

[11] Z. Wang, A. Scaglione, and R. J. Thomas, "The Node Degree Distribution in Power Grid and Its Topology Robustness under Random and Selective Node Removals," in *2010 IEEE International Conference on Communications Workshops*, 2010, pp. 1–5.

[12] D. J. Watts and S. H. Strogatz, "Collective dynamics of 'small-world' networks," *Nature*, vol. 393, no. 6684, pp. 440–442, Jun. 1998.

[13] Z. Wang and R. J. Thomas, "On Bus Type Assignments in Random Topology Power Grid Models," in *2015 48th*

*Hawaii International Conference on System Sciences*, 2015, pp. 2671–2679.

[14] Z. Wang, S. H. Elyas, and R. J. Thomas, "Generating Synthetic Electric Power System Data with Accurate Electric Topology and Parameters," in *UPEC 2016*, 2016.

[15] S. H. Elyas and Z. Wang, "A Multi-objective Optimization Algorithm for Bus Type Assignments in Random Topology Power Grid Model," in *2016 49th Hawaii International Conference on System Sciences (HICSS)*, 2016, pp. 2446–2455.

[16] Z. Wang, S. H. Elyas, and R. J. Thomas, "A novel measure to characterize bus type assignments of realistic power grids," in *2015 IEEE Eindhoven PowerTech*, 2015, pp. 1–6.

[17] S. H. Elyas and Z. Wang, "Improved Synthetic Power Grid Modeling with Correlated Bus Type Assignments," *IEEE Trans. Power Syst.*, pp. 1–1, 2016.

[18] S. H. Elyas, Z. Wang, and R. J. Thomas, "On Statistical Size and Placement of Generation and Load for Synthetic Grid Modeling," in *The 10th Bulk Power Systems Dynamics and Control Symposium - (IREP 2017)*, 2017.

[19] Z. Wang and S. H. Elyas, "On the Scaling Property of Power Grids," in *Hawaii International Conference On System Sciences HICSS*, 2017.

[20] S. H. Elyas and Z. Wang, "Statistical analysis of transmission line capacities in electric power grids," in *2016 IEEE Power & Energy Society Innovative Smart Grid Technologies Conference (ISGT)*, 2016, pp. 1–5.

[21] A. B. Birchfield, T. Xu, K. M. Gegner, K. S. Shetye, and T. J. Overbye, "Grid Structural Characteristics as Validation Criteria for Synthetic Networks," *IEEE Trans. Power Syst.*, pp. 1–1, 2016.

[22] K. M. Gegner, A. B. Birchfield, Ti Xu, K. S. Shetye, and T. J. Overbye, "A methodology for the creation of geographically realistic synthetic power flow models," in *2016 IEEE Power and Energy Conference at Illinois (PECI)*, 2016, pp. 1–6.

[23] A. B. Birchfield, K. M. Gegner, T. Xu, K. S. Shetye, and T. J. Overbye, "Statistical Considerations in the Creation of Realistic Synthetic Power Grids for Geomagnetic Disturbance Studies," *IEEE Trans. Power Syst.*, pp. 1–1, 2016.

[24] T. Xu, A. B. Birchfield, K. S. Shetye, and T. J. Overbye, "Creation of Synthetic Electric Grid Models for Transient Stability Studies," in *The 10th Bulk Power Systems Dynamics and Control Symposium (IREP 2017)*, 2017.

[25] M. E. Raoufat, K. Tomsovic, and S. M. Djouadi, "Virtual Actuators for Wide-Area Damping Control of Power Systems," *IEEE Trans. Power Syst.*, vol. 31, no. 6, pp. 4703–4711, Nov. 2016.

[26] M. E. Raoufat, K. Tomsovic, and S. M. Djouadi, "Power system supplementary damping controllers in the presence of saturation," in *2017 IEEE Power and Energy Conference at Illinois (PECI)*, 2017, pp. 1–6.

[27] M. E. Raoufat, K. Tomsovic, and S. M. Djouadi, "Dynamic Control Allocation for Damping of Inter-area Oscillations," *IEEE Trans. Power Syst.*, pp. 1–1, 2017.

[28] A. Birchfield *et al.*, "A Metric-Based Validation Process to Assess the Realism of Synthetic Power Grids," *Energies*, vol. 10, no. 8, p. 1233, Aug. 2017.

[29] H. Sangrody, M. Sarailoo, N. Zhou, N. Tran, M. Motalleb, and E. Foruzan, "Weather forecasting error in solar energy forecasting," *IET Renew. Power Gener.*, Jun. 2017.

[30] A. Shokrollahi, H. Sangrody, M. Motalleb, M. Rezaeiahari, E. Foruzan, and F. Hassanzadeh, "Reliability Assessment of Distribution System Using Fuzzy Logic for Modelling of Transformer and Line Uncertainties," in *49th North American Power Symposium (NAPS)*, 2017, pp. 1–6.

[31] A. Shahsavari *et al.*, "A Data-Driven Analysis of Capacitor Bank Operation at a Distribution Feeder Using Micro-PMU Data," in *2017 IEEE Power & Energy Society Innovative Smart Grid Technologies Conference (ISGT)*, 2017, pp. 1–5.

[32] M. H. Athari and Z. Wang, "Statistically Characterizing the Electrical Parameters of the Grid Transformers and Transmission Lines," in *The 10th Bulk Power Systems Dynamics and Control Symposium – (IREP 2017)*, 2017.

[33] S. Kullback and R. Leibler, "On information and sufficiency," *Ann. Math. Stat.*, vol. 22, no. 1, pp. 79–86, 1951.

[34] D. MacKay, *Information theory, inference and learning algorithms*. Cambridge university press, 2003.